\documentclass[12pt]{article}

\usepackage{times}

\newcommand{\DAG}{DAG}

\begin{document}

\noindent
Peter M. \textsc{Aronow} and Fredrik \textsc{S\"avje} \\
Yale University\\
115 Prospect Street, New Haven, CT 06520, USA\\
peter.aronow@yale.edu and fredrik.savje@yale.edu\\
\medskip

\noindent{\Large The Book of Why: The New Science of Cause and Effect}
\begin{quote}
Judea \textsc{Pearl} and Dana \textsc{Mackenzie}.
New York: Basic Books, 2018. ISBN 978-0-465-09761-6.
ix+432 pp.
\end{quote}

\noindent Book review published as: Aronow, Peter M. and  Fredrik S\"avje (2020), ``The Book of Why: The New Science of Cause and Effect." \emph{Journal of the American Statistical Association}, 115: 482--485.

\medskip

\hrulefill

\medskip

\thispagestyle{empty}
\raggedright\baselineskip=18pt\parindent=2em\parskip=0pt

Judea Pearl is a giant in the field of causal inference, whose many contributions, including the discovery of the \textit{d}-separation criterion, have been immeasurably valuable.
He, along with science writer Dana Mackenzie, has written an important book that relates Pearl's work to a broad audience and makes an argument for its place in the scientific canon.

The book recounts the history of the Causal Revolution.
The reader is told that causal inference was in a sorry state of affairs for most of the twentieth century.
The scientific community was unable to tackle even the most basic causal inquiry, with grave consequences.
For example, when discussing how scientists could not reach an agreement about whether smoking causes lung cancer, the book notes that ``millions of lives were lost or shortened because scientists did not have an adequate language or methodology for answering causal questions'' (p.\ 19).
However, during this age of darkness, a small light of hope was burning in the form of Sewall Wright and a few other brave men.
The light was the spark of the Causal Revolution in the 1990s.
The revolution consisted of the introduction of a graphical causal representation in the form of directed acyclic graphs (\DAG{}s), and a set of associated tools, by Pearl and his coauthors.
Scientists were finally given the language and methodology they needed to conduct serious causal investigations, and apart from a few pockets of resistance, the scientific community rejoiced:
\begin{quote}
There is now an almost universal consensus, at least among epidemiologists, philosophers and social scientists, that (1) confounding needs and has a causal solution, and (2) causal diagrams provide a complete and systematic way of finding that solution. (p.\  141)
\end{quote}

As an autobiographical and expositional text for those working in the field, the book is both informative and entertaining.
We are, however, concerned that the presentation will mislead readers whose first acquaintance with the subject is this book.
The goals of this review are twofold: to highlight instances where naive readers, especially policy-makers, might be led to unrealistically optimistic conclusions; and to discuss alternative models that are overlooked in Pearl's account of the causal revolution.

\subsubsection*{Unfounded optimism about causal models}

The book paints a picture of a field that has come to its conclusion.
Causal inference, for most intents and purposes, is solved.
The optimism is appealing: the world is just waiting for scientists to uncover its mysteries.
A scientific revolution will follow Pearl's causal revolution, at least among those that adopt his language and methodology.

We believe this optimism is unfounded.
The central problem that scientists face, especially in the social sciences, is not how to express or analyze causal models but how to pick one that is valid or at least reasonable.
The book does not claim to solve this problem.
For example, discussing Henry Niles' critique of Sewall Wright's research, Pearl writes:
\begin{quote}
Many people still make Niles' mistake of thinking that the goal of causal analysis is to prove that $X$ is a cause of $Y$ or else to find the cause of $Y$ from scratch.
That is the problem of causal discovery \ldots\
In contrast, the focus of Wright's research, as well as this book, is representing plausible causal knowledge in some mathematical language, combining it with empirical data and answering causal queries that are of practical value. Wright understood from the very beginning that causal discovery was much more difficult and perhaps impossible. (pp.\ 79--80)
\end{quote}
The topic of the book is causal analysis, not causal discovery.
In Pearl's model and calculus, the underlying causal structure is assumed to be known.
The issue is that scientists often disagree on this structure.
Pearl's approach may help clarify exactly where the disagreement lies, but it will not provide adjudication.

The smoking-lung cancer debate, recounted in Chapter~5, is a good illustration.
Following Pearl's recipe, the scientific community in the 1950s should have represented the current consensus about the plausible causes of cancer (and all other relevant aspects of the causal nexus) in a \DAG{}.
After applying Pearl's calculus, the causal relationships of interest could have been investigated empirically, and the debate would have been resolved.
The recipe fails, however, already on the first step.
As captured in the exchange between Ronald Fisher and Jerome Cornfield, the central question of the debate was which causal model was the plausible one.
Fisher held it possible (or plausible) that a gene was causing both smoking and lung cancer, and Cornfield disagreed.
Diagrams~5.1 and~5.2 in the book (p.\ 176) provide a graphical representation of their positions.
The disagreement was, however, not founded in a misunderstanding, so the clarification provided by the \DAG{}s would have been of little use.

The heart of the problem is that large parts of the scientific community share Pearl's skepticism about the prospects of causal discovery.
In our experience, the modes of inquiry that adjudicate debates avoid the problem altogether.
Cornfield's sensitivity analysis, which played an important role in resolving the smoking and lung cancer debate, is one such approach.
Pearl acknowledges its value, but the approach is outside of his dichotomy of causal analysis and causal discovery.
The causal structure (e.g., as encoded in a \DAG{}) is not presumed to be known here, nor is that structure the target of our inferences, as in causal discovery.
Instead, sensitivity analyses can remain somewhat agnostic about the underlying structure.
This agnosticism is one reason they are useful; sensitivity analyses can adjudicate debates because scientists can agree on their validity without reaching full agreement over what constitutes plausible causal knowledge.

The book's treatment of randomized controlled trials (RCTs) also fails to demonstrate an appreciation for this central problem.
Pearl writes:
\begin{quote}
Once we have understood why RCTs work, there is no need to put them on a pedestal and treat them as the gold standard of causal analysis, which all other methods should emulate.
Quite the opposite: we will see that the so-called gold standard in fact derives its legitimacy from more basic principles. (p.\ 140)
\end{quote}
For the purposes of causal analysis in Pearl's model, there is no distinction between a randomized experiment and any similarly unconfounded treatment assignment.
However, this neglects that the successful implementation of an experiment ensures the assumption of unconfoundedness in a manner that can survive scrutiny from even the most determined skeptic.
We can move from a metaphysical discussion about the correct causal model to a practical discussion about the experimental protocol and whether it was followed.
Experiments allow us to be largely agnostic about the causal structure.

The problem with these more agnostic modes of inference is their limited applicability.
As Pearl notes, we can do less without a rich causal model.
No one doubts the usefulness of Pearl's framework in situations where it can be applied.
What naive readers might miss is that these situations are rare, particularly in the social sciences.
And in the cases where reasonable consensus about the causal structure can be reached, Pearl's calculus will often tell us that it is not possible to draw inferences.
 For example, as noted in the book, Fisher's claim that genetic disposition confounds the smoking-lung cancer relationship was correct. A graph alone cannot encode that this confounding is weak, and Pearl's calculus would have told us that progress was not possible until these genes were identified and measured.

These realizations are sobering: contrary to the impression given to readers, causal inquiry cannot be reduced to a mathematical exercise nor automatized.
Causal inference is possible, but it is a daunting task best served by modesty and humility.

\subsubsection*{Pluralism in causal inference}

Pearl's self-described ``Whig history'' of causal inference is selective and narrow.
Besides his own contributions, the book focuses on the contributions of his intellectual ancestors.
Antagonists are occasionally brought on the stage, but only for the purpose of being proven wrong.
Readers will easily be under the impression that the field has seen a slow but inevitable progression towards enlightenment, despite misguided resistance from the establishment, culminating with Pearl as a singular figure.

This account is misleading.
No consensus, not even an emerging one, exists about the superiority of \DAG{}s.
Causal inference has its roots in many disciplines, and several conceptual frameworks and methodological approaches exist and thrive.
Pearl reduces this pluralism to ``cultural resistance'' (p.\ 394).
Scientists who resist \DAG{}s are, however, not stubborn monks using their quills to defend a last stand against the printing press.
The pluralism is instead a reflection of the range of challenges they face.
The reason Pearl's model is not used more widely is that many scientists do not find it useful.

An extensive survey of the field is beyond the scope of this review.%
\footnote{If such a survey were to be written, it could address the two most notable omissions in the book.
The first is the contributions of a large group of scholars who were central to the development of modern causal inference.
A partial list of this group is: Angrist, Ashenfelter, Campbell, Card, Heckman, Imbens, Manski, Murphy, Robins, and Rosenbaum.
The second omission is estimation from data, which the book has a tendency to trivialize.
Readers are not made aware of the fundamental difficulties in this enterprise.
Under Pearl's model, it is impossible to estimate causal effects well without additional statistical assumptions.
Statistical theory in this setting remains an active area of research.}
We will instead provide a few examples of alternative causal models and explain why scientists might prefer them.
These approaches differ from Pearl's model in that they impose different amounts or different types of structure on the causal problem.

Robins (1986) developed a causal model that is closely related to Pearl's model and predates it.
Both models are nonparametric structural equation models.
The sole distinction is that Pearl's model invokes additional independence assumptions.%
\footnote{Robins (1986) also introduces notation and a calculus for causal effects, called g-notation and g-computation, that have direct parallels with Pearl's \textit{do}-operator and \textit{do}-calculus.
In particular, Robins' g-notation includes Pearl's \textit{do}-operator as a special case.}
These assumptions make his model more powerful, but they do not have testable implications.
Pearl is quick to point the first part while neglecting the second.
For example, Robins' model does not allow for identification of natural mediation effects, and Pearl notes that thanks to his model the ``age-old quest for a mediation mechanism has been reduced to an algebraic exercise'' (p. 20).
The reason Robins did not make these assumptions is not because of a lack of imagination.
He was reluctant to do so because they are too strong for the applications he has in mind, and because they cannot be verified even through experimentation.
These are exactly the considerations scientists face.
The additional assumptions do make the model more powerful, but they change its interpretation, and they make the analysis less robust.
Pearl notes in passing that one must feel comfortable with his assumptions, but he neglects to tell the reader what they entail, being quick to instead give them a ``green light'' on mediation analysis (p. 334).

The next model is the framework our own methodological work has focused on, namely the design-based approach to causal inference.
Drawing its origins from the work of both Jerzy Neyman and Fisher, scientists using this approach consider random allocation of treatment as the sole source of stochasticity.
The framework is, in this sense, agnostic: it does not presume the existence of any stable laws of nature nor any infinite super-population from which we can draw new observations.
Inferences are motivated primarily by knowledge of the experimental design, rather than by unverifiable statistical assumptions.
As we noted above, however, agnosticism has costs.
A design-based model focuses on a circumscribed task.
Its purpose is to investigate causal effects of a well-defined intervention for a fixed population of units in one particular setting.

There are situations where scientists need more expressive models to solve their empirical problems.
For example, parametric or semiparametric structural equation models can encode not only the existence (or non-existence) of causal relationships but also \emph{how} different variables are related.
Scientists often find this useful when their substantive knowledge or theories suggest certain functional forms.
An economist might, for instance, be comfortable assuming that an increase in income will not cause a reduction in total consumption.
The usual trade-off applies, of course, and more elaborate models may introduce conceptual ambiguities and lack of robustness.

Pearl does not hide the fact that his calculus cannot exploit information about characteristics of causal relations, but he does not explain to what degree this limits its usefulness.
An illustrative example lies with Pearl's discussion of the local average treatment effect interpretation of the instrumental variable method.
The approach relies on a monotonicity assumption, which states that a causal effect is in the same direction for all units in the population.
The assumption concerns the characteristics of an effect rather than its existence, so it cannot be encoded in a \DAG{}.
In Pearl's words:
\begin{quote}
In \textit{do}-calculus we make no assumptions whatsoever regarding the nature of the functions in the causal model.
But if we can justify an assumption like monotonicity or linearity on scientific grounds, then a more special-purpose tool like instrumental variables estimation is worth considering. (p. 257)
\end{quote}
The statement is correct but misleading.
A casual reader would be under the impression that these assumptions and the associated ``special-purpose tools'' are on the fringes of causal inference.
On the contrary, instrumental variable methods are immensely popular in the social sciences.
So are regression discontinuity and difference-in-differences designs, which are other methods relying on functional form assumptions (continuity and additivity, respectively).
These methods are omitted from Pearl's account of the causal revolution.
If readers were made aware of their existence and popularity, they might question whether ``causal diagrams provide a complete and systematic way of finding a solution [to confounding].''

\begin{flushright}\def\baselinestretch{1}
Peter M. \textsc{Aronow} and Fredrik \textsc{S\"avje}\\
\emph{Yale University}
\end{flushright}

\subsubsection*{References}

\noindent Robins, James (1986), ``A new approach to causal inference in mortality studies with a sustained exposure period---application to control of the healthy worker survivor effect,'' \emph{Mathematical Modelling}, 7, 1393--1512.

\end{document}